\documentclass[aps,prd,a4paper,preprintnumbers,twocolumn,floatfix,showpacs,nofootinbib,usenatbib]{revtex4-1}
\usepackage{float,graphicx,multirow,bm,color,amsmath,amssymb,hyperref}
\hypersetup{
    colorlinks=true,
    linkcolor=green,
    citecolor=blue,
}

\begin{document}

\title{$f(T)$ gravity and local Lorentz invariance}
\author{Baojiu~Li,{}$^{1,2}$ Thomas~P.~Sotiriou,{}$^{1}$ and John~D.~Barrow{}$^1$}
\address{$^1$DAMTP, Centre for Mathematical Sciences, University of Cambridge, Cambridge CB3 0WA, UK\\ 
$^2$Kavli Institute for Cosmology Cambridge, Madingley Road, Cambridge CB3 0HA, UK}
\date{\today}

\begin{abstract}
We show that in theories of generalised teleparallel gravity, whose
Lagrangians are algebraic functions of the usual teleparallel Lagrangian, the action and the
field equations are \emph{not} invariant under local Lorentz
transformations. We also argue that these theories appear to have extra
degrees of freedom with respect to general relativity. The usual teleparallel Lagrangian, which has been extensively studied and leads to a theory dynamically equivalent to general relativity, is an exception. Both of these facts
appear to have been overlooked in the recent literature on $f(T)$ gravity, but are crucial for assessing the viability
of these theories as alternative explanations for the acceleration of the universe.
\end{abstract}

\pacs{04.50.Kd, 04.20.Fy, 11.30.Cp, 95.36.+x, 98.80.-k}
\maketitle

\affiliation{Kavli Institute for Cosmology Cambridge, Madingley Road,
Cambridge CB3 0HA, UK} \affiliation{DAMTP, Centre for Mathematical Sciences,
University of Cambridge, Cambridge CB3 0WA, UK}

\affiliation{DAMTP, Centre for Mathematical Sciences, University of
Cambridge, Cambridge CB3 0WA, UK}

\affiliation{DAMTP, Centre for Mathematical Sciences, University of
Cambridge, Cambridge CB3 0WA, UK}



Teleparallel gravity \cite{Einstein, Aldrovandi} is a gravity theory which
uses the curvature-free Weitzenbock connection \cite{Weitzenbock} to define
the covariant derivative, instead of the conventional torsionless
Levi-Civita connection of general relativity, and attempts to describe the
effects of gravitation in terms of torsion instead of curvature. In its
simplest form it is equivalent to general relativity (GR) but has a
different physical interpretation \cite{Aldrovandi}. Motivated by attempts
to explain the observed acceleration of the universe in a natural way, there
has been a great deal of recent interest in a generalisation of this theory
in which the Lagrangian is an arbitrary algebraic function $f$ of the Lagrangian of teleparallel gravity $T$. This is in direct analogy to creating $f(R)$ gravity theories as a
generalisation of GR (see Ref.~\cite{Sotiriou:2008rp} for a review). This, so-called $f(T)$ gravity theory, has cosmological solutions
which could provide alternative explanations for the acceleration of the
universe  \cite{Bengochea:2009, Yu:2010a, Myrzakulov:2010a, Tsyba:2010,
Linder:2010, Yi:2010b, Kazuharu:2010a, Kazuharu:2010b, Myrzakulov:2010b,
Yu:2010c, Karami:2010, Dent:2010}.  
The field equations for the $f(T)$ gravity have been claimed to be
very different from those for $f(R)$ gravity, as they are second order
rather than fourth order. This has been considered as an indication that the
theory may be the more interesting relative of GR.

Here we will look further into the symmetries and dynamics of $f(T)$
gravity. Our main findings will be that such theories are not locally
Lorentz invariant and appear to harbour extra degrees of freedom not present
in GR. Remarkably, both of these features have been overlooked in the
literature.

Let us briefly introduce teleparallel gravity and its $f(T)$ generalisation.
Our dynamical variables are the vierbein or tetrad fields, $\mathbf{h}%
_{a}\left( x^{\mu }\right) $, which form an orthonormal basis for the
tangent space at each point of the manifold with spacetime coordinates $%
x^{\mu }$. Latin indices label tangent space coordinates while Greek indices
label spacetime coordinates. All indices run from $0$ to $3$. Clearly $%
\mathbf{h}_{a}\left( x^{\mu }\right) $ is a vector in tangent space, and can
be described in a coordinate basis by its components $h_{a}^{\mu }$. So, $%
h_{a}^{\mu }$ is also a vector in spacetime.  

The spacetime metric, $g_{\mu \nu }$, is given by 
\begin{equation}
\label{eq:metric}
g_{\mu \nu }=\eta _{ab}h_{\mu }^{a}h_{\nu }^{b}\,,
\end{equation}%
where $\eta _{ab}=\mathrm{diag}(1,-1,-1,-1)$ is the Minkowski metric for the
tangent space. It follows that 
\begin{equation}
h_{a}^{\mu }h_{\nu }^{a}\ =\ \delta _{\nu }^{\mu },\ \ \ h_{a}^{\mu }h_{\mu
}^{b}\ =\ \delta _{a}^{b},
\end{equation}%
where Einstein's summation convention has been used. GR uses the Levi-Civita
connection 
\begin{equation}
\Gamma _{\mu \nu }^{\lambda }\equiv \frac{1}{2}g^{\lambda \sigma }\left(
g_{\sigma \mu ,\nu }+g_{\sigma \nu ,\mu }-g_{\mu \nu ,\sigma }\right),
\end{equation}%
in which commas denotes partial derivatives. This leads to nonzero spacetime
curvature but zero torsion. In contrast, teleparallel gravity uses the
Weitzenbock connection $\tilde{\Gamma}_{\ \mu \nu }^{\lambda }$(tilded to
distinguish it from $\Gamma _{\mu \nu }^{\lambda }$), 
\begin{equation}
\tilde{\Gamma}_{\ \mu \nu }^{\lambda }\equiv h_{b}^{\lambda }\partial _{\nu
}h_{\mu }^{b}\ =\ -h_{\mu }^{b}\partial _{\nu }h_{b}^{\lambda }
\end{equation}%
which leads to zero curvature but nonzero torsion. The torsion tensor reads 
\begin{equation}
\label{torsion}
T_{\phantom{a}\mu \nu }^{\lambda }\equiv \tilde{\Gamma}_{\ \nu \mu
}^{\lambda }-\tilde{\Gamma}_{\ \mu \nu }^{\lambda }\ =\ h_{b}^{\lambda
}\left( \partial _{\mu }h_{\nu }^{b}-\partial _{\nu }h_{\mu }^{b}\right) .
\end{equation}%
The difference between the Levi-Civita and Weitzenbock connections, which
are not tensors, is a spacetime tensor, and is known as the contorsion
tensor: 
\begin{eqnarray}
K_{\phantom{a}\mu \nu }^{\rho } &\equiv &\tilde{\Gamma}_{\ \mu \nu }^{\rho
}-\Gamma _{\mu \nu }^{\rho }\ =\ \frac{1}{2}\left( T_{\mu \ \nu }^{\ \rho
}+T_{\nu \ \mu }^{\ \rho }-T_{\ \mu \nu }^{\rho }\right)   \notag
\label{eq:contorsion} \\
&=&h_{a}^{\rho }\nabla _{\nu }h_{\mu }^{a},
\end{eqnarray}%
where $\nabla _{\nu }$ denotes the metric covariant derivative.

If one further defines the tensor $S^{\rho \mu \nu }$ as 
\begin{equation}
S^{\rho \mu \nu }\equiv K^{\mu \nu \rho }-g^{\rho \nu }T_{\ \ \ \sigma
}^{\sigma \mu }+g^{\rho \mu }T_{\ \ \ \sigma }^{\sigma \nu },
\end{equation}%
then the teleparallel lagrangian density is given by 
\begin{equation}
\mathcal{L}_{T}\equiv \frac{h}{16\pi G}T\ \equiv \ \frac{h}{32\pi G}S^{\rho
\mu \nu }T_{\rho \mu \nu },
\end{equation}%
in which $h=\sqrt{-g}$ is the determinant of $h_{a}^{\lambda }$ and $g$ is
the determinant of the metric $g_{\mu \nu }$, $G$ is the gravitational
constant and 
\begin{eqnarray}
T &\equiv &\frac{1}{2}S^{\rho \mu \nu }T_{\rho \mu \nu }=-S^{\rho \mu \nu
}K_{\rho \mu \nu }  \notag  \label{eq:T} \\
&=&\frac{1}{4}T^{\rho \mu \nu }T_{\rho \mu \nu }+\frac{1}{2}T^{\rho \mu \nu
}T_{\nu \mu \rho }-T_{\rho \mu }^{\ \ \rho }T_{\ \ \ \nu }^{\nu \mu }.
\end{eqnarray}%
Variation with respect to the tetrad $h_{\lambda }^{a}$ after adding a
matter Lagrangian density $\mathcal{L}_{m}$ leads to the field equations 
\begin{equation}
h^{-1}\partial _{\sigma }\left( hh_{a}^{\rho }S_{\rho }^{\ \lambda \sigma
}\right) -h_{a}^{\sigma }S^{\mu \nu \lambda }T_{\mu \nu \sigma }+\frac{1}{2}%
h_{a}^{\lambda }T=8\pi G\,\Theta _{a}^{\lambda },  \label{eq:field_eqn_tel}
\end{equation}%
where $\Theta _{a}^{\lambda }\equiv h^{-1}\delta \mathcal{L}_{m}/\delta h_{%
\phantom{a}\lambda }^{a}$. The usual stress-energy tensor is given in
terms of $\Theta _{a}^{\lambda }$ as $\Theta ^{\mu \nu }=\eta ^{ab}\Theta
_{a}^{\nu }h_{b}^{\mu }$.

The $f(T)$ gravity theory generalises $T$ in the lagrangian density to an
arbitrary function of $T$: 
\begin{equation}
\label{ftaction}
\mathcal{L}_{T}\rightarrow \mathcal{L}\ =\ \frac{h}{16\pi G}f(T).
\end{equation}%
The derivation of field equations is very similar to that described above
for teleparallel gravity. They are 
\begin{eqnarray}\label{eq:field_eqn}
f_T\left[h^{-1}\partial_\sigma \left(hh_a^\rho S_{\rho}^{\ \lambda\sigma}\right) - h^\sigma_a S^{\mu\nu\lambda}T_{\mu\nu\sigma}\right] \\
+ f_{TT}h_a^\rho S_{\rho}^{\ \lambda\sigma}\partial_\xi T + \frac{1}{2}h^\lambda_a f(T) &=& 8\pi G\,\Theta_a^{\lambda}\,,\nonumber
\end{eqnarray}
where $f_{T}\equiv \partial f(T)/\partial T$ and $f_{TT}\equiv \partial
^{2}f(T)/\partial T^{2}$. Clearly, for $f(T)=T,$ Eq.~(\ref{eq:field_eqn})
reduces to Eq.~(\ref{eq:field_eqn_tel}).

We now move on to consider the symmetries of the action and the dynamical
content of the field equations. When working in terms of tetrads and making
explicit reference to a tangent space, two invariance principles should hold 
\cite{weinbergbook}: the action should be a generally covariant scalar, and
so invariant under the infinitesimal coordinate transformations $x^{\mu
}\rightarrow x^{\mu }+\epsilon ^{\mu }(x)$;\ and if special relativity is to
be recovered in locally inertial frames, the action must also be invariant
under local (position-dependent) Lorentz transformations (\emph{i.e.}~we
should be able to redefine the locally inertial coordinate systems at each
point). Let us check if these properties hold for $f(T)$ gravity.

We start with the matter action, which in the literature is assumed to
couple to the tetrad so as to couple effectively only to the metric. In this
case the matter action is, as usual, both a generally covariant scalar and a
Lorentz scalar\footnote{Dropping this assumption for the matter coupling would lead to Lorentz
violations in the matter sector.}. It is worth considering  the consequences
of these assumptions for the matter action as an explicit example.

We denote an infinitesimal Lorentz transformation as $\Lambda _{\
b}^{a}(x^{\mu })=\delta _{\ b}^{a}+\omega _{\ b}^{a}(x^{\mu })$ with $%
|\omega _{\ b}^{a}|\ll 1$ and $\omega _{ab}=\omega _{\lbrack ab]}$. Square brackets denote anti-symmetrisation and parentheses symmetrisation. As the
vierbein $h_{a}^{\mu }$ is a Lorentz vector in index $a$, it changes by $%
\delta h_{a}^{\mu }=\omega _{a}^{\ b}h_{b}^{\mu }$ under this Lorentz
transformation, where we have suppressed the dependence on $x^{\mu }$ for
simplicity. The matter action 
\begin{equation}
\mathcal{S}_{m}=\int d^{4}x\mathcal{L}_{m}
\end{equation}%
is then changed by \cite{weinbergbook, Aldrovandi} 
\begin{equation}
\delta \mathcal{S}_{m}=\int \Theta _{\ \mu }^{a}h\delta h_{a}^{\ \mu
}d^{4}x\ =\ \eta ^{bc}\int \Theta _{\ \mu }^{a}h\omega _{ab}h_{c}^{\ \mu
}d^{4}x\,.  \label{eq:invariance_lorentz}
\end{equation}%
 $\omega _{ab}$ is an arbitrary antisymmetric (Lorentz) tensor, and 
\begin{equation}
\eta ^{bc}\Theta _{\ \mu }^{a}h_{c}^{\ \mu }\ =\ \eta ^{ac}\Theta _{\ \mu
}^{b}h_{c}^{\ \mu }\ \Leftrightarrow \ \Theta ^{\beta \alpha }\ =\ \Theta
^{\alpha \beta },
\end{equation}%
so we see that $\delta \mathcal{S}_{m}=0$ yields 
\begin{equation}
\Theta ^{\beta \alpha }=\Theta ^{\alpha \beta }.  \label{symTh}
\end{equation}%
In other words, if $\mathcal{S}_{m}$ is invariant under local Lorentz
transformations, then $\Theta _{\mu \nu }$ is symmetric, and vice versa.

Consider now the fact that the matter action is invariant under the
infinitesimal coordinate transformation $x^{\mu }\rightarrow x^{\mu
}+\epsilon ^{\mu }(x)$ where $|\epsilon ^{\mu }|\ll 1$. Under this
transformation the vierbein changes by $\delta h_{a}^{\mu }(x)=h_{a}^{\nu }\epsilon _{,\nu }^{\mu }-h_{a,\lambda}^{\mu }\epsilon ^{\lambda }$ \cite{weinbergbook} 
and the invariance of $\mathcal{S}_{m}$ yields 
\begin{eqnarray}
\label{eq:invariance_spacetime}
0 
&=&
\int d^{4}x\epsilon ^{\lambda }\left[ \partial _{\nu }\left( h\Theta _{\
\lambda }^{a}h_{a}^{\nu }\right) +h\Theta _{\ \mu }^{a}h_{a,\lambda }^{\mu }%
\right] 
\end{eqnarray}%
where we have dropped a total derivative. Now $\epsilon ^{\lambda }$ is an
arbitrary spacetime vector, so we must have 
\begin{eqnarray}
0 &=&\partial _{\nu }\left( h\Theta _{\ \lambda }^{a}h_{a}^{\nu }\right)
+h\Theta _{\ \mu }^{a}h_{a,\lambda }^{\mu }  \notag  \label{thcon} \\
&=&h\nabla ^{\nu }\Theta _{\nu \lambda }+h\Theta ^{\alpha \nu }K_{\alpha \nu
\lambda }.
\end{eqnarray}%
Given that $K_{(\mu \nu )\rho }=0$ and using Eq.~(\ref{symTh}), we get 
\begin{equation}
\nabla ^{\nu }\Theta _{\nu \lambda }=0.
\end{equation}%
Clearly, if $\Theta _{\mu \nu }$ were not symmetric, \emph{i.e.}~if the matter action 
were not invariant under local Lorentz transformations, then $\Theta _{\mu \nu }$ would not be divergence-free either.

We now move to the gravitational sector. As already mentioned, $T_{%
\phantom{a}\mu \nu }^{\lambda }$ behaves like a tensor under spacetime
coordinate transformations (the antisymmetry of the last two indices allows
us to promote the partial derivatives to covariant ones). The last line of
Eq.~(\ref{eq:contorsion}) demonstrates that $K_{\phantom{a}\mu \nu }^{\rho }$
is also a spacetime tensor. Consequently, $S^{\rho \mu \nu }$ is also a
spacetime tensor and $T$ is a generally covariant scalar. Hence any action
constructed with $\mathcal{L}_{T}$ or $\mathcal{L}$ is generally covariant
and invariant under the infinitesimal coordinate transformation $x^{\mu
}\rightarrow x^{\mu }+\epsilon ^{\mu }(x)$.

Some more algebra is needed to check whether such actions are also local
Lorentz scalars. From the relation between $\Gamma _{\beta \gamma }^{\alpha }
$ and $\tilde{\Gamma}_{\beta \gamma }^{\alpha }$ given in Eq.~(\ref%
{eq:contorsion}), and the fact that the curvature tensor associated with the
Weitzenbock connection $\tilde{\Gamma}_{\beta \gamma }^{\alpha }$ vanishes,
we can write the Riemann tensor for the connection $\Gamma _{\beta \gamma
}^{\alpha }$ as \cite{Aldrovandi} 
\begin{eqnarray}
R_{\phantom{a}\mu \lambda \nu }^{\rho } &=&\partial _{\lambda }\Gamma _{%
\phantom{a}\mu \nu }^{\rho }-\partial _{\nu }\Gamma _{\phantom{a}\mu \lambda
}^{\rho }+\Gamma _{\phantom{a}\sigma \lambda }^{\rho }\Gamma _{\phantom{a}%
\mu \nu }^{\sigma }-\Gamma _{\phantom{a}\sigma \nu }^{\rho }\Gamma _{%
\phantom{a}\mu \lambda }^{\sigma } \\
&=&\nabla _{\nu }K_{\phantom{a}\mu \lambda }^{\rho }-\nabla _{\lambda }K_{%
\phantom{a}\mu \nu }^{\rho }+K_{\phantom{a}\sigma \nu }^{\rho }K_{\phantom{a}%
\mu \lambda }^{\sigma }-K_{\phantom{a}\sigma \lambda }^{\rho }K_{\phantom{a}%
\mu \nu }^{\sigma }.  \notag
\end{eqnarray}%
The corresponding Ricci tensor is then 
\begin{eqnarray}
R_{\mu \nu } &=&\nabla _{\nu }K_{\phantom{a}\mu \rho }^{\rho }-\nabla _{\rho
}K_{\phantom{a}\mu \nu }^{\rho }+K_{\phantom{a}\sigma \nu }^{\rho }K_{%
\phantom{a}\mu \rho }^{\sigma }-K_{\phantom{a}\sigma \rho }^{\rho }K_{%
\phantom\mu \nu }^{\sigma }  \notag  \label{eq:Ricci} \\
&=&-\nabla ^{\rho }S_{\nu \rho \mu }-g_{\mu \nu }\nabla ^{\rho }T_{%
\phantom{a}\rho \sigma }^{\sigma }-S_{\phantom{ab}\mu }^{\rho \sigma
}K_{\sigma \rho \nu },
\end{eqnarray}%
and the Ricci scalar 
\begin{equation}
\label{ricciscalar}
R=-T-2\nabla ^{\mu }\left( T_{\phantom{a}\mu \nu }^{\nu }\right) .
\end{equation}%
The relations 
\begin{eqnarray}
&&K^{(\alpha \beta )\gamma }=T^{\alpha (\beta \gamma )}=S^{\alpha (\beta
\gamma )}=0,  \notag \\
&&S_{\phantom{a}\rho \mu }^{\mu }=2K_{\phantom{a}\rho \mu }^{\mu }\ =-2T_{%
\phantom{a}\rho \mu }^{\mu },
\end{eqnarray}%
and Eq.~(\ref{eq:T}) were used in deriving Eqs.~(\ref{eq:Ricci}) and (\ref%
{ricciscalar}).

Eq.~({\ref{ricciscalar}) is very useful, as it shows that $T$ and $R$ differ
only by a total divergence. This immediately implies that $\mathcal{L}_{T}$
is completely equivalent to the Einstein--Hilbert lagrangian density, as the total
divergence can be neglected inside an integral, and teleparallel gravity is
equivalent to GR. We will see this below at the level of the field equations
as well. For the moment, let us focus on a different feature. $R$ is a
generally covariant scalar and also a local Lorentz scalar as it can be
expressed in terms of the metric and without any reference to the tetrad.
Now $\nabla ^{\mu }\left( T_{\phantom{a}\mu \nu }^{\nu }\right) $ is also a
generally covariant scalar, as $T_{\phantom{a}\mu \nu }^{\lambda }$ is a
spacetime tensor. Thus, as argued above, $T$ is a generally covariant
scalar. However, $\nabla ^{\mu }\left( T_{\phantom{a}\mu \nu }^{\nu }\right) 
$ is \emph{not} a local Lorentz scalar: as one can easily check, it is not
invariant under a local Lorentz transformation. Consequently, $T$ is \emph{%
not} a local Lorentz scalar either.

 This has been pointed out already in the literature of standard teleparallel gravity (see Ref.~\cite{Aldrovandi} and references therein), {\em i.e.}~when the action considered is constructed simply with $\mathcal{L}_{T}$, as well as in studies of more general theories where the action is constructed with the Weitzenbock connection and is quadratic in the torsion tensor \cite{Hehl:1978yt,Hayashi:1979qx,MuellerHoissen:1983vc,Cheng:1988zg,Blagojevic:2000qs, FR}. The former case is very special as the resulting theory is still locally
Lorentz invariant. The reason is that the Lorentz breaking term is a
total divergence. Therefore, the apparent lack of local Lorentz symmetry at the level of the action appears to be of little importance in
teleparallel gravity, \emph{i.e.}~when the Lagrangian is just $T$.

 However, the situation is quite different for the $f(T)$ generalisation of teleparallel gravity. It is clear that if $T$ is not a local
Lorentz scalar then $f(T)$ cannot be either. Moreover, $f(T)$
cannot be split into two parts with one a local Lorentz scalar and the other
a total divergence. This implies that actions of the form given in Eq.~(\ref{ftaction})
are not locally Lorentz invariant. So, $f(T)$ generalizations are not special as the standard teleparallel gravity where $f(T)=T$, but instead behave like the more generic theories where a general action constructed with a Weitzenbock connection is considered.  \footnote{Even though this was mentioned already in Ref.~\cite{Ferraro:2007} for a specific action which falls under the general $f(T)$ class, the implications of the lack of local Lorentz symmetry were not fully spelt out.}

To get a better understanding of this, we can verify what was said above
also at the level of the field equations. Contracting with $h_{\nu }^{a}$
and using Eqs.~(\ref{eq:Ricci}) and (\ref{ricciscalar}), after some
algebra we can bring Eq.~(\ref{eq:field_eqn}) into the form 
\begin{eqnarray}
H_{\mu \nu } &\equiv &f_{T}G_{\mu \nu }+\frac{1}{2}g_{\mu \nu }\left[
f(T)-f_{T}T\right] +f_{TT}S_{\nu \mu \rho }\nabla ^{\rho }T  \notag
\label{eq:modified_einstein_eqn} \\
&=&8\pi G\Theta _{\mu \nu },
\end{eqnarray}%
where $G_{\mu \nu }$ is the Einstein tensor. When $f(T)=T$, GR is recovered,
which verifies the claim that teleparallel gravity and GR are equivalent. In
this case the field equations are clearly covariant and the theory is also
local Lorentz invariant. In the more general case with $f(T)\neq T$,
however, this is not the case. Even though all terms in Eq.~(\ref%
{eq:modified_einstein_eqn}) are covariant, the last two terms in the first
line are not local Lorentz invariant. Hence the field equations are not
invariant under a local Lorentz transformation. $\ $

Local Lorentz invariance would mean that we can only determine the tetrad up
to a local Lorentz transformation; that is, only 10 of the 16 components of
the tetrad would be independent and fixing the rest would simply be a gauge
choice. Lack of Lorentz  invariance implies that the field equations must
determine these 6 components as well, leading to a system of 16 equations
instead of 10. This is indeed the case: notice that $H_{\mu \nu }$ is not
symmetric, but $\Theta _{\mu \nu }$ is, because matter is assumed to  couple
only to the metric (see above). Therefore, we can split Eq.~(\ref%
{eq:modified_einstein_eqn}) in the following way 
\begin{eqnarray}
H_{(\mu \nu )} &=&8\pi G\Theta _{\mu \nu },  \label{H1} \\\label{H2}
H_{[\mu \nu ]} &=&0,
\end{eqnarray}%
which forms a system of 16 component equations. As in GR, we can do away
with 4 of these equations by using the usual spacetime gauge symmetry, but 
there still remain 6 more equations. Note also that since the action and the
field equations are covariant, and matter is assumed to couple only to the
metric, $H_{\mu \nu }$ does satisfy a generalised contracted Bianchi
identity. This means that the zero divergence of $\Theta _{\mu \nu }$
imposes no further constraints. This can be easily argued at the level of
the action in analogy with the treatment of $\Theta _{\mu \nu }$ above
(modulo the symmetry), but it can also be demonstrated by a direct
calculation. Using the definition of $H_{\mu \nu }$ that 
\begin{equation}
\nabla ^{\mu }H_{\mu \nu }=f_{TT}\left[ R_{\mu \nu }+g_{\mu \nu }\nabla
^{\sigma }T_{\phantom{a}\sigma \rho }^{\rho }+\nabla ^{\sigma }S_{\nu \sigma
\mu }\right] \nabla ^{\mu }T,
\end{equation}%
and Eq.~(\ref{eq:Ricci}), one gets 
\begin{eqnarray}
&&\nabla ^{\mu }H_{\mu \nu }+f_{TT}\nabla ^{\mu }T\,S_{\phantom{ab}\mu
}^{\rho \sigma }K_{\sigma \rho \nu }  \notag \\
&=&\nabla ^{\mu }H_{\mu \nu }+H^{\sigma \rho }K_{\sigma \rho \nu }=0.
\end{eqnarray}%
For the first equality we have used the fact the $K_{(\sigma \rho )\nu }=0$%
. This equation is in direct agreement with the analogous equation for $%
\Theta _{\mu \nu }$, Eq.~(\ref{thcon}). If we now use Eq.~(\ref{H2}), and  $%
K_{(\sigma \rho )\nu }=0$ again, then we get 
\begin{equation}
\nabla ^{\mu }H_{\mu \nu }=0.
\end{equation}%
Therefore, on shell, $H_{\mu \nu }$ satisfies a generalized contracted
Bianchi identity, as expected from our symmetry analyses above. This is
typically the case for covariant theories with extra degrees of freedom
non-minimally coupled to gravity, \emph{e.g.}~scalar-tensor\ gravity
theories. Indeed, the theory appears to propagate more degrees of freedom,
as is consistent with the lack of symmetry. Eqs. (\ref{H1}) and (\ref{H2})
are second-order differential equations but they are expected to harbour
more degrees of freedom than the two graviton polarizations of GR, contrary
to what has been implied in the literature. Also, the fact that the field
equations are second order does not mean that extra excitations will
necessarily be healthy. For instance, a wrong sign could lead to ghosts or
classically unstable modes. The dynamics of the extra degrees of freedom of  
$f(T)$ gravity certainly deserves further investigation.

Lack of local Lorentz symmetry implies that there is no freedom to fix any
of the components of the tetrad. They must all be determined by the field
equations. Now, suppose that we want to impose a metric ansatz based on
specific spacetime symmetry assumptions. Does this imply a certain ansatz
for the tetrad? The answer is, only partially. Eq.~(\ref{eq:metric})
provides only 10 algebraic relations between the 10 independent metric
components and the 16 independent tetrad components. Were the theory local
Lorentz invariant, one would be able to fix the remaining 6 tetrad
components. In absence of the symmetry this is not an option and they need
to be determined by the field equations.

For instance, assuming a spatially flat Friedmann--Lema\^{\i}%
tre--Robertson--Walker line element, 
\begin{equation}
ds^{2}=dt^{2}-a(t)^{2}(dx^{2}+dy^{2}+dz^{2})
\end{equation}
does not uniquely lead to the tetrad choice 
\begin{equation}
h_{\mu }^{a}=\mathrm{diag}(1,a(t),a(t),a(t)),
\end{equation}%
as is very commonly assumed in the literature. There is simply not enough
freedom to make this assumption and one would need to resort to the field
equations and explicitly show, not only the consistency, but also the
uniqueness of this specific choice.

To summarise, we have studied the symmetries and the dynamics of $f(T)$
theories of gravity. We have shown that, even though they are covariant,
such theories are not local Lorentz invariant, with the exception of the $%
f(T)=T$ case, which have been extensively studied in the literature and is equivalent to GR. This fact has several consequences.
First, it is expected to lead to strong preferred-frame effects which should
in turn be crucial for the viability of the theory. This casts serious doubt
on whether such theories can provide interesting alternatives to GR. Note
that even though matter will not `feel' the preferred frame effects because
it is only coupled to the metric, these effects still leave an observational
signature in gravitational experiments, as in the case of Einstein-aether
theory \cite{Jacobson:2008aj}. Another consequence is that the lack of
symmetry implies the presence of more degrees of freedom. Indeed, there
appear to be 6 more dynamical equations than in GR. Even though all
equations are second order in derivatives, this is not enough to guarantee
that the extra excitations will be well behaved. The lack of Lorentz
symmetry also presents a serious computational complication because there is
no freedom to gauge fix tetrad components. 

We hope that this analysis will prompt a search for a deeper understanding
of the dynamics of $f(T)$ gravity, the presence of extra degrees of freedom
in these theories, and their cosmological behaviour. There also needs to be
a thorough study on the observational consequences of local Lorentz symmetry
violations. We hope to address these issues in future work.

Before closing let us point out that it is rather trivial to modify $f(T)$ theory in order to make it manifestly Lorentz invariant. If the partial derivative is replace by a Lorentz covariant derivative (see Ref.~\cite{weinbergbook}) in the definition of $T^\lambda_{\phantom{a}\mu\nu}$, Eq.~(\ref{torsion}), and then one defines a quantity  $\bar{T}$ in the same way as $T$ is defined here, $\bar{T}$ or $f(\bar{T})$ will be manifestly locally Lorentz invariant, see also Ref.~\cite{Arcos:2010gi}. Note, however, that even though such a theory will reduce to $f(T)$ gravity in some local Lorentz frames (those for which the Lorentz covariant derivative becomes a partial derivative), it will generically have different dynamics. It is, therefore, a different theory, which might deserve further investigation.

\emph{Acknowledgements:} We thank R.~Ferraro and P.~G.~Pereira for helpful discussions.
B.~Li is supported by Queens' College, University of Cambridge
and the Science and Technology Facilities Council (\texttt{STFC}) of the UK.
T.~P.~Sotiriou is supported by a Marie Curie Fellowship.

\end{document}